\documentclass[aps,prb,preprint,showpacs,superscriptaddress]{revtex4}
\usepackage{bm, amsmath, amssymb}
\usepackage{graphicx}

\newcommand{\figwidth}{0.47\textwidth}

\begin{document}

\title{Magnetic vortex as a ground state  for micron-scale
antiferromagnetic samples }

\author{E.~G. Galkina}
\affiliation{Institute of Physics, 03028 Kiev, Ukraine}
\affiliation{RIKEN Advanced Science Institute, Wako-shi, Saitama,
351-0198, Japan}

\author{A.~Yu. Galkin}
\affiliation{Institute of Metal Physics, 03142 Kiev, Ukraine }

\author{B.~A. Ivanov}
\email{bivanov@i.com.ua} \affiliation{Institute of Magnetism, 03142,
Kiev, Ukraine} \affiliation{RIKEN Advanced Science Institute,
Wako-shi, Saitama, 351-0198, Japan}

\author{Franco Nori}
\affiliation{RIKEN Advanced Science Institute, Wako-shi, Saitama,
351-0198, Japan} \affiliation{Department of Physics, The University
of Michigan, Ann Arbor, MI 48109-1040, USA, }
\date{\today}

\begin{abstract}
Here we consider micron-sized  samples with any axisymmetric body
shape and made with a canted antiferromagnet, like hematite or
iron borate. We find that its  ground state can be a magnetic
vortex with a topologically non-trivial distribution of the
sublattice magnetization $\vec{l}$ and planar coreless vortex-like
structure for the net magnetization $\vec{M}$. For
antiferromagnetic samples in the vortex state, in addition to
low-frequency modes, we find high-frequency modes with frequencies
over the range of hundreds of gigahertz, including a mode
localized in a region of radius $\sim$ 30--40 nm near the vortex
core.
\end{abstract}

\maketitle

\section{Introduction}
The magnetic properties of submicron ferromagnetic samples shaped
as circular cylinders (magnetic dots) are attracting considerable
attention mainly due to their potential applications (see, e.g.,
Ref.~\onlinecite{Skomski03}). Circular dots possess an equilibrium
magnetic configuration which corresponds to a vortex structure
just above the single domain length scale, with radius $R >
R_{\mathrm{crit}} \sim$ 100--200 nm. The ferromagnetic vortex
state consists of an in-plane flux-closure magnetization
distribution and a central core with radius $\sim$ 20--30 nm,
magnetized perpendicular to the dot plane. Reducing the
magnetostatic energy comes at the cost of a large exchange energy
near the vortex core, as well as the magnetostatic energy caused
by the core magnetization. Due to spatial quantization, the magnon
modes for dots have a discrete spectrum. The possibility to
control the localization and interference of magnons spawned
\emph{magnonics}, trying to use these modes for developing a new
generation of microwave devices with submicron active elements
(see, e.g., Ref.~\onlinecite{magnonics}).

These samples also provide an ideal experimental system for
studying static and especially dynamic properties of relatively
simple topologically non-trivial magnetic structures which are
fundamentally interesting objects in different research areas of
physics. The investigation of the non-uniform states of ordered
media with non-trivial topology of the order parameter can be
considered as one of the most impressive achievements of modern
condensed matter physics (see, e.g.,
Ref.~\onlinecite{VolovikBook}), and field theory (see, e.g.,
Ref.~\onlinecite{two}). For example, vortices (topological
solitons) appear in many systems with continuously degenerate
ground states, whose properties are determined by some phase-like
variable $\phi $, including superconductors (see, e.g.,
Ref.~\onlinecite{deGennes}), quantum liquids (helium-II, and
different phases of superfluid $^3$He, see, e.g.,
Ref.~\onlinecite{VolovikBook}), dilute Bose-Einstein
condensates,~\cite{PitBEC} and also different models of magnets:
ferromagnets and antiferromagnets,~\cite{BICh,KIK90,BarIvKhalat}
and spin nematics.~\cite{IvKolPRB03,IvKolKhymPRL08} The
contribution of topological excitations (vortices and vortex
pairs) to the thermodynamics and response functions of a
two-dimensional ordered media are well known. At low temperatures,
vortices are bound into pairs, forming a Berezinskii phase with
absence of long-range order, but with quasi-long-range order. The
unbinding of the vortex pairs at high enough temperatures, $T >
T_{\mathrm{BKT}}$, leads to the Berezinskii-Kosterlitz-Thouless
phase transition.  In particular, the translational motion of
vortices leads to a central peak in the dynamic correlation
functions, which has been observed experimentally; see
Refs.~\onlinecite{WieslerZabel,MertensRev00,BarIvKhalat} and
references therein.

Physical systems with a  topological defect as a ground state are
of special interest. The role of vortices in rotating superfluid
systems is known.\cite{VolovikBook} Additional examples include
single-connected samples of the A-phase of superfluid $^3$He,
where the true energy minimum is a non-trivial state (bujoom) with
a surface vortex-like singularity.\cite{VolovikBook} Moreover,
non-uniform states can also appear in small magnetic samples
having high enough surface anisotropy.\cite{DimWysin,SurfAniz}
Among these examples, magnets are interesting because samples
bearing vortices can be prepared of micrometer and sub-micrometer
size, and vortices can be present at high enough temperatures,
including room temperature. Furthermore, their static and dynamic
properties can be observed using  different techniques.(e.g.,
Refs.~\onlinecite{Skomski03,Antos+Rev,GuslRev})
 \subsection{Summary of results}
  All previous studies of magnetic vortices caused by
magnetic dipole interactions were carried out on magnetic samples
made with soft ferromagnets with large magnetization $M_s $, like
permalloy, with $4\pi M_s \sim 1$ T. In this work, we show that
vortices can be a ground state for micron-scale samples with axial
symmetry, but now made with another important class of magnetic
materials: antiferromagnets with easy-plane anisotropy and
Dzyaloshinskii--Moriya interaction (DMI). Some antiferromagnets
possess a small, but non--zero, net magnetization caused by a weak
non-collinearity of the sublattices (sublattice canting)
originated from the DMI. As we show here, the formation of a
vortex in an AFM sample leads to a more effective minimization of
the  non-local magnetic dipole interaction, compared to the case
with ferromagnetic dots. Typical antiferromagnets include hematite
$\alpha$-$\mbox{Fe}_2 \mbox{O}_3 $, iron borate $\mathrm{FeBO}_3
$, and orthoferrites.~\cite{BICh} These materials exhibit magnetic
ordering at high temperatures. Moreover, they have unique physical
properties: orthoferrites and iron borate are transparent in the
optical range and have a strong Faraday effect, and the
magnetoelastic coupling is quite high in hematite and iron
borate.~\cite{BICh} For antiferromagnets, typical frequencies of
spin oscillations have values in a wide region, from gigahertz to
terahertz, which can be excited by different techniques (not only
the standard one using of ac-magnetic fields, but with optical or
acoustic methods as well).  Spin oscillations can also be
triggered  by ultra-short laser
pulses.~\cite{Kimel,kimel+09,Kalash,Satoh+} For antiferromagnetic
(AFM) samples in the vortex state,  here we derive  a rich
spectrum of discrete magnon modes with frequencies from
sub-gigahertz to hundreds of gigahertz, including a mode localized
near the vortex core. These results should help  to extend the
frequency range of magnonic  microwave devices till sub-terahertz
region.

\section{Model}
For antiferromagnets, the exchange interaction between neighboring
spins facilitates an antiparallel spin orientation, which leads to
a structure with two antiparallel magnetic sub-lattices, $\vec
{M}_1 $ and $\vec {M}_2 $, $\vert \vec {M}_1 \vert =\vert \vec
{M}_2 \vert =M_0 $. To describe the structure of this
antiferromagnet, it is convenient to introduce irreducible
combinations of the vectors $\vec {M}_1 $ and $\vec {M}_2 $, the
net  magnetization $\vec {M}=\vec {M}_1 +\vec {M}_2 =2M_0 \vec
{m}$, and the sublattice magnetization vector $\vec {l}=(\vec
{M}_1 -\vec {M}_2 )/2M_0 $. The vectors $\vec {m}$ and $\vec {l}$
are subject to the constraints: $(\vec {m} \cdot \vec {l})=0$ and
$ \vec {m}^2+\vec {l}^2=1$.  As $\vert \vec {m}\vert \ll 1$, the
vector $\vec {l}$ could be considered as a unit vector. The mutual
orientation of the sublattices is determined by the sum of the
energy of uniform exchange $W_\mathrm{ex}$ and the DMI energy,
$W_{\mathrm{DM}}$,
$$W_\mathrm{ex}=H_\mathrm{ex} M_0 \vec {m}^2 \,,  \  \  \
W_{\mathrm{DM}}=2M_0 H_\mathrm{D} [\vec {d} \cdot (\vec {m}\times
\vec {l})]\,,$$
    where the unit vector $\vec {d}$ is directed along
the symmetry axis of the magnet. Here we will not discuss the role
of external magnetic fields. The parameters $H_{\mathrm{ex}} \sim
3\cdot 10^2$--$10^3 $ T and $H_\mathrm{D} \sim 10$ T are the
exchange field and DMI field, respectively. Using the energy
$(W_{\mathrm{ex}}+W_{\mathrm{DM}})$ and the dynamical equations
for $\vec {M}_1 $ and $\vec {M}_2 $, one can find,\cite{BICh}
\begin{equation}
\label{eq5} \vec {M}=M_{\mathrm{DM}} \left(\vec {d}\times \vec
{l}\right)+ \frac{2M_0}{\gamma H_{\mathrm{ex}}}\left(\vec {l} \times
\frac{\partial \vec {l}}{\partial t}\right) ,\, M_{\mathrm{DM}} =
\frac{2H_{\mathrm{D}} M_0 }{H_{\mathrm{ex}}} \, ,
\end{equation}
where $\gamma $ is the gyromagnetic ratio.  The first term gives
the static value of the antiferromagnet net magnetization
$M_{\mathrm{DM}} $, comprising a small parameter, $H_\mathrm{D}
/H_{\mathrm{ex}} \sim 10^{-2}$, and the second term describes the
dynamic canting of sublattices; see Ref.~\onlinecite{BICh} for
details. Note that $M_{\mathrm{DM}} $ is much smaller than either
$M_0 $ or the value of $M_s$ for typical ferromagnets. However,
the role of the magnetostatic energy caused by $M_{\mathrm{DM}} $
could be essential, and could lead to the appearance of a domain
structure for antiferromagnets,~\cite{book,BICh} and the magnetic
dipolar stabilization of the long-range magnetic order for the 2D
case.\cite{IvTartPRL} We will show that for the formation of
equilibrium vortices, antiferromagnets have some advantages
compared with soft ferromagnets.

The static and especially dynamic properties of antiferromagnets
are essentially different from the ones of ferromagnets. The spin
dynamics of an antiferromagnet can be described using the
so-called sigma-model ($\sigma$-model), a dynamical equation only
for the vector $\vec{l}$ (see, e.g.,  Ref.~\onlinecite{BICh}).  In
this approach, the magnetization $\vec M = \vec M_1+\vec
M_2=2M_0\vec m $ is a slave variable and can be written in terms
of the vector $\vec l$ and its time derivative, see
Eq.~(\ref{eq5}). Within the $\sigma$-model, the equation for the
normalized (unit) antiferromagnetic vector $\vec l = \vec L /|
\vec L|$ can be written through the variation of the Lagrangian
$\mathcal{L}[\vec l]$
\begin{equation}
\label{eqL} \mathcal{L} = \frac{A }{2 c^2} \int
{\left(\frac{\partial \vec l}{\partial t}\right)^2}d^3\!x -
W\left[\vec l \right] \,,
\end{equation}
where $A$ is the non-uniform exchange constant, $c = \gamma
\sqrt{AH_\mathrm{ex}/M_0}$ is the characteristic speed describing
the AFM spin dynamics, $W[\vec {l}] $ is the functional describing
the static energy of the AFM. It is convenient to present the energy
functional in the form $W\bigl[\vec l \bigr] = W_0\bigl[\vec l
\bigr]+ W_m $,
\begin{eqnarray}\label{eq3}
\nonumber
  W_0\bigl[\vec l \bigr] &=& \frac{1}{2}\int {[A(\nabla \vec {l})^2+K\cdot l_z^2 ]}
d^3 x \, , \\
  W_m &=& -\frac{1}{2} \int {\vec {M}\vec {H}_m } d^3\!x \, ,
\end{eqnarray}
where the first term $W_0[\vec {l}]$ determines the local model of
an AFM, including the energy of non-uniform exchange and the
easy-plane anisotropy energy through only the vector $\vec {l}$,
$K>0$ is the anisotropy constant, and the $xy-$plane is the easy
plane for spins. The second (non-local) term $W_m $ is the
magnetic dipole energy, $\vec {H}_m $ is the demagnetization field
caused by the AFM magnetization \eqref{eq5}, the field $\vec {H}_m
$ is determined by the magnetostatic equation,
  $$ \mathrm{div}(\vec {H}_m+4\pi \vec M)=0,\,
   \mathrm{curl}\,\vec {H}_m =0\,, $$
$\vec M= 2M_0\vec m$, with the standard boundary conditions: the
continuity of the normal component of $(\vec {H}_m+4\pi \vec M)$
and the tangential component of $\vec {H}_m$, on the border of the
sample. Thus, sources of $\vec {H}_m $ can be considered as formal
``magnetic charges'', where volume charges equal to $\mathrm{div}
\vec {M}$ and surface charges equal $-\vec {M}\cdot \vec {n}$,
where $ \vec {n}$ is the unit vector normal to the border ( see
the monographs~\onlinecite{book},~\onlinecite{SW} for general
considerations and Refs.~\onlinecite{IvZasAPL,IvZasPRL05} for
application to vortices). As mentioned above, the presence of
surface anisotropy with the constant $K_{\mathrm{surf}}$ can
contribute to the energy of a micron-scale sample leading to
non-uniform states.\cite{DimWysin,SurfAniz} But this contribution
is essential for a high enough anisotropy $K_{\mathrm{surf}}\gg K$
and quite small samples, and we will neglect this contribution,
and concentrate on an alternative source of non-uniform states,
namely, the magnetic dipole interaction.
\subsection{Dynamic sigma-model}
The variation of the total Lagrangian \eqref{eqL} gives a dynamic
$\sigma$-model equation, where the static spin distribution for an
uniaxial antiferromagnet is determined by the minimization of an
energy functional of the form of Eq.~\eqref{eq3}. It is useful to
start with a variation of the local energy $W_0[\vec {l}]$ only,
that gives a general two-dimensional (2D) vortex solution for the
vectors $\vec {l}$, $\vec {M}$ of the form
\begin{eqnarray}\nonumber
\label{eq1} \vec {l}&=&\vec {e}_z \cos \theta +\sin \theta [\vec
{e}_x \cos (\chi +\varphi _0 )+\vec {e}_y \sin (\chi +\varphi _0 )]\,,\\
\vec {M}&=&M_\mathrm{DM}\cdot \sin \theta [\vec {e}_y \cos (\chi
+\varphi _0 )-\vec {e}_x \sin (\chi +\varphi _0 )],
\end{eqnarray}
where  $\theta=\theta (r)$, $r$ and $\chi $ are polar coordinates
in an easy-plane of the magnet, the vector $\vec {e}_z $ is the
hard axis, and the value of $\varphi _0 $ is arbitrary. The
function $\theta (r)$ is determined by the ordinary differential
equation
\begin{equation}\label{eqTheta0}
\frac{d^2\theta}{dx^2}+\frac{1}{x}\frac{d \theta}{dx}=\sin\theta
\cos\theta \left(\frac{1}{x^2}-1 \right),\, x=\frac{r}{l_0} \,,
\end{equation}
and exponentially tends to $\pi /2$ for $r \gg l_0 $, with
characteristic size $l_0 =\sqrt {A/K} $. Also, in the center of
the vortex (at $r$ = 0), the value of $\sin \theta (r=0)=0$, see
Fig.~1.

In the region of the vortex core, $\vec l$ deviates from the
easy-plane, and the anisotropy energy increases. The state
(\ref{eq1}) is non-uniform,  increasing  the exchange energy.
Therefore, for the local easy-plane model, the appearance of a
vortex costs some energy, i.e., the vortex corresponds to excited
states of antiferromagnets. Also, vortex excitations are important
for describing of thermodynamics of 2D
antiferromagnets.~\cite{AFMvort91} As we will show in the next
section,  considering  the dipolar energy $W_m[\vec l]$, a vortex
can be the ground state of a circular magnetic sample made with
canted antiferromagnets.

\begin{figure}[htbp]
\includegraphics[width=\figwidth]{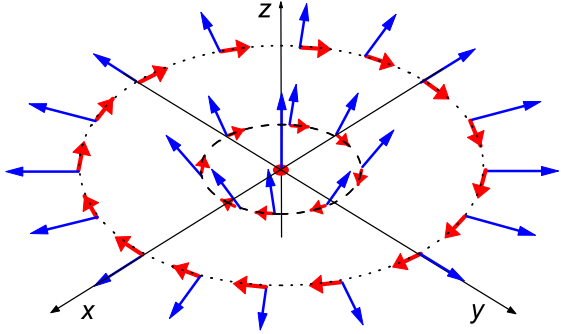}
\caption{\label{fig1} (Color online) Schematic structure of the
AFM vortex for $\cos \theta (r=0)=1$ and the special value of the
arbitrary parameter $\varphi _0 =0$. The vectors $\vec {l}$ (thin
blue arrows in the radial direction) and $\vec {M}$ (thick
tangential red arrows; shown not in scale) are depicted for the
area of the core (dashed circle) and far from it ( larger dotted
circle). The red dot at the origin indicates the value $\vec
{M}=0$ for the state with $\vec {l}=\vec e_z$ perpendicular to the
plane.}
\end{figure}

\section{Energy of the vortex and uniform states}
For small samples made with canted antiferromagnets, the energy
loss caused by a vortex within the local model $W=W_0[\vec l]$ can
be compensated by the energy of the magnetic dipole interaction.
To explain this, note that for a uniform distribution state, the
contribution $W_m $ unavoidably results in a loss of the system
energy, which is proportional to the volume $V$ of the
sample.~\cite{SW,book} The energy of the uniform state could be
estimated as
\begin{equation}\label{Ehomog}
E^{(\mathrm{homog})}=2\pi NM_{DM}^2 V=2\pi NM_0^2 (2H_\mathrm{D}
/H_e )^2V,
\end{equation}
where $N$ is the effective demagnetizing factor in the direction
perpendicular to the sample axis.~\cite{SW,book}

For  magnetic samples other than an ellipsoid (for example, for a
cylinder with a finite aspect ratio $R/L$), the distribution of
the field $\vec H_m$ is non-uniform, leading to a non-topological
quasi-uniform state known as a flower state.~\cite{Ross} But a
detailed numerical analysis
shows,~\cite{UsovFMM,HolPerevorot,GuslNovosVortStat,KravchukFTT07}
in all the stability region  of the quasi-uniform state where
Eq.~\eqref{Ehomog} is valid, the corresponding value of $N$ is
practically the same as for the uniform magnetization, and the
numerical
results~\cite{UsovFMM,HolPerevorot,GuslNovosVortStat,KravchukFTT07}
obtained within this approximation are in  good agreement with
experiments.~\cite{Ross}

For the static vortex state \eqref{eq1} with a chosen value of $\sin
\varphi _0 =0$, with $\vec M \propto (\vec {d} \times \vec {l})$,
one can find
\begin{equation} \label{eq7}
\vec {M}=\sigma  M_\mathrm{DM}\cdot \sin \theta (-\vec {e}_x \sin
\chi +\vec {e}_y \cos \chi ),
\end{equation}
where $\sigma =\cos \varphi _0 =\pm 1$. A unique property of the
state \eqref{eq7} is that it can also \emph{exactly} minimize the
energy of the magnetic dipole interaction $W_m$, giving $\vec
{H}_m =0$ in the overall space. Indeed, the projection of $\vec
{M}$ on the lateral surface of any axisymmetric body with its
symmetry axis parallel to the $z$-axis equals to zero. Also
$\mathrm{div} \vec {M} = 0$ for the distribution given by
Eq.~\eqref{eq7}. Moreover, due to the symmetry of the DMI, $(\vec
{M} \cdot \vec {d})=0$, the distribution of the magnetization
$\vec {M}$ (\ref{eq7}) is \emph{purely planar} (in contrast to
$\vec {l}$) and the out-of-plane component of $\vec {M}$ is
absent. In the vicinity of the vortex core, the length of the
vector $\vec {M}$ decreases, turning to zero in the vortex center,
see Fig.~1. Such feature is well known for domain walls in some
orthoferrites.~\cite{BICh} Thus, the AFM vortex is the unique spin
configuration which does not create a demagnetization field in a
singly-connected body (for ferromagnet with
$|\vec{M}|=\mathrm{const}$, a configuration with $\vec H_m \equiv
0$ can be possible only for magnetic rings having the topology of
a torus).
\subsection{Comparing the energies of the vortex and uniform states for antiferromagnets}

Let us now compare the energies of the vortex state and the
uniform state for AFM samples  shaped as a cylinder with height
$L$ and radius $R$. For the vortex state, $\vec {H}_m =0$ over the
the sample volume, and the vortex energy is determined by the
simple formula,~\cite{BICh}
\begin{equation}\label{Evort}
E_{\mathrm{v}}=\pi AL\ln \left( {\rho \frac{R}{l_0 }} \right),
\end{equation}
where $\rho \approx 4.1$ is a numerical factor. For long cylinders
with $L \gg R$, the value of  $N\simeq 1/2$ and the vortex state
becomes favorable if the radius $R$ exceeds some critical value
$R_\mathrm{crit} $,
\begin{eqnarray} \nonumber
\label{eq9} R\ge R_{\mathrm{crit}} &=& 2 l_{\mathrm{dip}} \sqrt
{\ln
\left(  \frac{l_{\mathrm{dip}}}{l_0} \right)}\,,  \\
l_{\mathrm{dip}}&=&\sqrt {\frac{A}{4\pi M_{\mathrm{DM}}^2 }}\,,
\end{eqnarray}
where $l_{\mathrm{dip}}$ determines the spatial scale
corresponding to the magnetic dipole interaction. Note that
$l_{\mathrm{dip}}$ comprises a large parameter $H_e /H_\mathrm{D}
\sim 30$--$100$, and $l_{\mathrm{dip}} \gg l_0$. In the case of a
thin disk, $ L \ll R$ and the demagnetization field energy
becomes~\cite{IvZasAPL,IvZasPRL05}
$$E^{(\mathrm{homog})}=2\pi RL^2M_{DM}^2 \ln (4R/L),$$
 and the vortex state is energetically favorable for
$RL \ge (RL)_{\mathrm{crit}} =2 l_{\mathrm{dip}}^2$.

For concrete estimates we take the parameters of iron borate,
$A=0.7\cdot 10^{-6}$ erg/cm, $K=4.9\cdot 10^{6}$ erg/cm$^3$ and
$4\pi M_{\mathrm{DM}} =120$ Oe. Then we obtain that $l_0 =3.8$ nm,
i.e., the core size is of the same order of magnitude as for a
typical ferromagnet (for permalloy $l_0 =4.8$ nm). The value
$l_{\mathrm{dip}} $ is essentially higher, e.g., for iron borate
$l_{\mathrm{dip}} =220$ nm. Combining these data one finds for the
long cylinder $R_{\mathrm{crit}}  =0.9$ $\mu$m. For a thin disk
sample the characteristic scale has submicron value: $\sqrt
{(RL)_{\mathrm{crit}} } =0.4$ $\mu$m. Similar estimates are
obtained for orthoferrites, and somewhat higher values for
hematite. Thus, despite the fact that the characteristic values
for the dipole length $l_{\mathrm{dip}} $ for a ferromagnet and
antiferromagnet differ hundredfold, the characteristic critical
sizes differ not so drastically (for permalloy $R_\mathrm{crit}
\sim$ 100-200 nm). This is caused by the aforementioned fact that
the magnetic field created by the vortex core is completely absent
for the AFM vortex. The situation here is common to that for
ferromagnetic nanorings, where the vortex core is absent. Despite
the fact that the vortex core size in ferromagnetic dots is rather
small, the core contribution to $W_m$ for ferromagnetic dots of
rather big radius $R\ge 0.5\;\mu $m is negligible, but it becomes
essential for small samples with $R$ close to the critical size.
Note as well that the vortex core magnetic field in the
ferromagnet destroys the purely 2D character of the distribution
of $\vec {M}$ in Eq.~(\ref{eq1}), and the core size changes over
the thickness of the sample. For an AFM vortex, the value of $\vec
{H}_m $ equals exactly zero, and a truly 2D distribution of $\vec
{l}$ and $\vec {M}$, independent of the coordinate $z$ along the
body axis, is possible.

\section{Magnon modes for vortex state antiferromagnetic samples}
The dynamics of $\vec l$  considered within the $\sigma$-model
approach differs from the dynamics of a ferromagnetic
magnetization described by the Landau--Lifshitz equation. The main
difference is that the $\sigma$-model equation contains a
dynamical term with a second-order time derivative of $\vec{l}$,
whereas the Landau--Lifshitz equation is first-order in time. For
this reason, for antiferromagnets, two magnon branches exist,
instead of one for ferromagnets.~\cite{BICh} For both AFM modes,
the elliptic polarization of the oscillations of $\vec M_1$ and
$\vec M_2$ is such that the oscillations of the vector $\vec l$
have a linear polarization.~\cite{BICh} For an easy-plane
antiferromagnet, these two branches are a low-frequency
quasi-ferromagnetic (QFM) branch and a high-frequency
quasi-antiferromagnetic (QAF) branch, respectively. QFM magnons
involve oscillations of the vectors $\vec l$ and $\vec M$ in the
easy-plane, with a weak deviation of $\vec M $ from the easy-plane
caused by the last term in \eqref{eq5}.  The second QAF branch
corresponds to the out-of-plane oscillations of $\vec l$ with the
dispersion law
$$\omega _{\mathrm{QAF}} (\vec {k}) = \sqrt {\omega _\mathrm{g}^2 +
c^2\vec {k}^2} , \ \omega _\mathrm{g} = \gamma \sqrt
{2H_{\mathrm{ex}} H_\mathrm{a} }, $$ where $\vec {k}$ is the magnon
wave vector. The gap of the QAF branch, $\omega _\mathrm{g}$,
contains a large value $H_{\mathrm{ex}}$ and attains hundreds of
GHz. Thus compared to ferromagnets, both the magnon frequency and
speed for AFM dynamics contain a large parameter
$$\sqrt{H_\mathrm{ex}/H_\mathrm{a}}\sim
30\mathrm{-}100,$$
 ($H_\mathrm{ex}$  and $H_\mathrm{a}$ are the exchange field and the
anisotropy field, respectively) which can be referred as the
\emph{exchange amplification} of the dynamical parameters of AFM.
The frequency $\omega_\mathrm{g}$ of AFM magnon modes reaches
hundreds of GHz, with values $\sim$170 GHz for hematite, 100-500
GHz for different orthoferrites, and 310 GHz for iron
borate.~\cite{data} Recent studies using ultra-short laser pulses
showed the possibility to excite spin oscillations of non--small
amplitude for orthoferrites~\cite{Kimel,kimel+09} and iron
borate.~\cite{Kalash} This technique can be also extended to other
antiferromagnets, including those without Dzyaloshinskii-Moriya
interaction.~\cite{Satoh+}

Since the magnon spectra of bulk ferromagnets and antiferromagnets
differ significantly, one can expect an essential difference for
the magnon modes of the vortex state for both AFM and
ferromagnetic samples. Let us briefly recall the properties of
normal modes for disk-shaped vortex state ferromagnetic samples
(ferromagnetic dots). For such dots, the presence of a discrete
spectrum of magnon modes, characterized by the principal number
(the number of nodes) $n$ and the azimuthal number $m$, is well
established.~\cite{GiovanniniExper,BuessExper,BuessPRB05,ZaspIvPark05}
This spectrum includes a single low-frequency mode of precessional
motion of a vortex core ($n = 0$, $m = 1)$ with a resonant
frequency in the subGHz region,~\cite{GusIv} a set of radially
symmetrical modes with $m = 0$,~\cite{mZero} and also a system of
slightly splinted doublets with azimuthal numbers $m = \pm \vert
m\vert $, with frequencies $\omega _{|m|, n} \neq \omega _{-|m|,
n} $, and $\omega _{|m|, n} - \omega _{-|m|, n} \ll \omega _{|m|,
n}$; see Refs.~\onlinecite{IvZasAPL,IvZasPRL05,GusSlav+Doubl}. The
same classification is valid for vortices for local easy-plane
ferromagnets.~\cite{ISchMW}  Wysin~\cite{GaryMass} demonstrated
the direct correspondence of the gyroscopic character of vortex
dynamics and doublet splinting.

For an AFM in the vortex state, small sample each of two magnon
branches, QFM and QAF, produce a set of discrete modes with given
$n$ and $m$; however, their properties are different compared to
that of a ferromagnetic dot. Below we present a general analysis
of small oscillations above the vortex ground state.

\subsection{General equations and mode symmetry. }
The dynamics of small deviations from the AFM static vortex
solution will be considered here for a thin circular sample (AFM
dot) only, where the $z$-dependence of the vectors $\vec l$ and
$\vec m$ can be neglected. It is convenient to introduce a local
set of orthogonal unit vectors $\vec e_1$, $\vec e_2$ and $\vec
e_3$, where $\vec e_3$ coincides with the local direction of the
unit vector $\vec l$ in the vortex, $\vec e_3 = \vec l(x,\,y)=\cos
\theta_0 \vec e_z+\sin \theta_0 \hat{\vec r}$, see
Fig.~\ref{fig1}, $\vec e_2 = -\vec e_x \sin \chi + \vec e_y \cos
\chi = \hat{\vec \chi}$, and $\vec e_1=(\vec e_2\times \vec e_3)$.
It is easy to see that the projection $(\vec l \cdot \vec
e_1)=\vartheta $ describes small deviations of $\theta $ from the
vortex ground state, $\theta =\theta _0 \left( r \right)+\vartheta
$, where $\theta _0$ is the solution of  Eq.~\eqref{eqTheta0},
describing a static vortex structure, and $\mu = (\vec l \cdot
\vec e_2)$ is the azimuthal component of $\vec l$. In linear
approximation, the equations for $\vartheta$ and $\mu$ become
 a set of coupled partial differential equations,
\begin{eqnarray}  \label{eqLin} \nonumber
\left[ \nabla_x^2-V_1(x) -\hat H_1 \right]\vartheta +
\frac{2\cos\theta_0}{x^2} \frac{\partial \mu}{\partial \chi}
&=&\frac{l_0^2}{c^2}\frac{\partial^2 \vartheta }{\partial t^2}\,,
\\
\left[ \nabla_x^2-V_2(x) -\hat H_2 \right]\mu -
\frac{2\cos\theta_0}{x^2} \frac{
\partial \vartheta}{\partial \chi} &=& \frac{l_0^2}{c^2}\frac{\partial^2 \mu}{
\partial t^2}\,,
\end{eqnarray}
where $x = r/l_0$, $\theta_0=\theta_0(x)$ is the solution of
Eq.~\eqref{eqTheta0}, and $\nabla_x = l_0\nabla$. The equations
are symmetric in $\vartheta$ and $\mu$ with local
Schr\"odinger-type differential operators in front, as well as
with non-local parts $\hat H_1$, $\hat H_2$. The ``potentials''
$V_1$, $ V_2$ in local Schr\"odinger-type operators are determined
by $W_0\bigl[\vec l\bigr]$, see Eq.~\eqref{eq3}; they have the
same form as for easy-plane magnets,~\cite{ISchMW,IvKolW}
\begin{equation}  \label{V12}
V_{1} = \left(\frac{1}{x^2}-1\right)\cos2\theta_0,\,V_{2} =
\left(\frac{1}{x^2} -1\right)\cos^2\theta_0 -
\left(\frac{d\theta_0}{dx}\right)^2\,.
\end{equation}
Note that the potentials in this Schr\"odinger-type operators are
not small, but localized near the vortex core. Non-local
magnetostatic effects, defined by magnetic dipole interactions,
are included in the integral operators, $\hat H_1$, $\hat H_2$.
For a ferromagnetic vortex, their form was determined and their
role was discussed in Refs.~\onlinecite{IvZasPRL05,BuessPRB05}.
Generally, for antiferromagnets the magnetization includes not
only terms proportional to the in-plane components of $\vec l$,
but also time derivatives of the vector $\vec l$. For this reason,
the structure of these operators presented though the vector $\vec
l$ is much more complicated than the corresponding structure for
ferromagnets. But the non-local contributions to \eqref{eqLin} are
essential only for in-plane modes with low frequencies $\omega \ll
\gamma H_{DM} $, and for this case the dynamical part of the
magnetization is negligible, as shown below. By means of this
approximation, one can demonstrate that the operators $\hat
H_{1,2}$, corresponding  to volume and to edge magnetic charges,
take the same form as for a ferromagnetic vortex, after replacing
$M_s \to M_{DM}$ and $M_z \to 0$. In particular, the angular
dependence of the eigenfunctions for these operators is the same
as for ferromagnetic vortices, namely, $\hat H_{1,2}\exp (im
\chi)=\Lambda_{|m|} \exp (im \chi)$, where the integer $m$
($m=0,\, \pm1\, \pm 2,\ \ldots$) is the azimuthal number.  Thus
for AFM vortices, even  considering  the non-local magnetic dipole
interaction with $\vec M \propto (\vec l \times \vec d)$, the
separation of the radial and azimuthal parts of the deviations is
possible, and the magnon modes are of the form  $\exp (im
\chi)f_m(r)$. This property is of  importance for the problem of
magnon modes above the AFM vortex ground state.

Thus, the static part of the equations \eqref{eqLin}, both local
and non-local terms, have the same form as for the well-studied
case of the ferromagnetic vortex, but the dynamical parts differ
strongly. This produces a crucial difference in the magnon modes
of these magnets. For ferromagnetic vortices, the magnon
eigenstates $\{\vartheta _m \,, \mu_m\}$ depend on $\chi$ and $t$
in combinations as $\sin(m\chi+\omega t)$ or $\cos(m\chi+\omega
t)$, whereas for a AFM vortex, a more general ansatz of the form
\begin{eqnarray} \label{ansatz} \nonumber
\vartheta_\alpha  &=&f_\alpha (r)(Ae^{im\chi} +Be^{-im\chi})\exp(i\omega _\alpha t)+\mathrm{c.c.}\, \\
\mu_\alpha  &=&ig_\alpha (r)(Ae^{im\chi}
-Be^{-im\chi})\exp(i\omega _\alpha t)+\mathrm{c.c.}\,
\end{eqnarray}
is appropriate.~\cite{IvKolW} Here $\alpha =(n,m)$ is a full set
of discrete numbers labelling the magnon eigenstates, and $n$ is
the nodal number. Substitution of this ansatz demonstrates, in
contrast to the case of a ferromagnetic vortex, the full
degeneracy of the frequency over the sign of $m$.\cite{IvKolW}  As
a change of sign in the number $m$ can also be interpreted as a
change of the sense of rotation of the eigenmode (change of sign
in the eigenfrequency $\omega$), physically we have the situation
of two independent oscillators rotating clock- and
counter-clockwise with the same frequency (which can also be
combined to give two linear oscillators in independent
directions). This degeneracy was clearly demonstrated by solving
the ordinary differential equations for $f_\alpha$ and $g_\alpha$
\eqref{eqLin}, as well as due to direct numerical simulations of
the magnon modes above an AFM vortex.\cite{IvKolW} Thus, the
absence of gyroscopical properties for the $\sigma$-model equation
is manifested in the fact that for a AFM vortex the modes with
azimuthal numbers $m = \vert m\vert $ è $m = - \vert m\vert $ are
degenerate, i.e., the splitting of doublets with $m = \pm \vert
m\vert $, typical for the ferromagnetic vortex, is completely
absent.~\cite{IvKolW,AFMvort96}

Note one more important difference from the ferromagnetic case:
for the AFM vortex the coupling of in-plane and out-of-plane
oscillations comes only from the term with $(\cos\theta_0)
(\partial /\partial \chi)$. This means, that (i) for any mode the
coupling vanishes exponentially at $x\to \infty$; (ii) the modes
of radially symmetric $(m=0)$ in-plane and out-of-plane
oscillations are completely uncoupled. Both properties will be
employed below for calculating  of the magnon frequencies.

\subsection{Collective variables for vortex core oscillations.}
First note that the equation \eqref{eqLin} for an infinite magnet
has a simple zero-frequency solution with $m=1$
 $$ \vartheta = (\vec a \cdot \nabla \theta _0)\,, \ \mu = \sin \theta _0
 (\vec a \cdot \nabla \phi _0),$$
 where $\vec a$ is an arbitrary vector, and
$  \theta _0$, $\phi _0$ describe the static vortex solution.
Indeed, this perturbation describes a displacement of the vortex
for a (small) vector $\vec a$. Such ``zero mode'' appears for any
soliton problem, reflecting an arbitrary choice of the soliton
(vortex) position.~\cite{two} For finite-size magnets, such modes
beget low-frequency modes corresponding to the motion of a vortex
core. Their analysis with the equations \eqref{eqLin} is quite
complicated, but it can be done within the approach based on the
scattering amplitude formalism, which is developed for the
Gross-Pitaevski equation~\cite{PitBEC} and local models for easy
plane magnets.~\cite{ISchMW} But there is an easier and more
convenient way to calculate the frequency of this mode based on a
collective variable approach. Here the collective-variable is the
vortex coordinate $\vec X$, which motion is described by a
characteristic dynamic equation.~\cite{ISchMW,GusIv}

Thus, for an AFM vortex, as well as for ferromagnets, one can
expect the appearance of a special mode of vortex core
oscillations. But the dynamical equations for the AFM vortex core
coordinate differ significantly from that for ferromagnets. The
$\sigma$-model equation contains a dynamical term with a second
time derivative of $\vec{l}$, combined with gradients of $\vec{l}$
in the Lorentz-invariant form $d^2\vec l / dt^2 - c^2\nabla ^2
\vec {l}$, whereas the Landau--Lifshitz equation is first-order in
time. The chosen speed $c = \gamma \sqrt{AH_\mathrm{ex}/M_0}$
plays roles of both the magnon speed and the speed limit of
solitons, it is only determined by the exchange interaction and
attains tens km/s; e.g., $c \simeq 1.4 \cdot 10^4$ m/s for iron
borate and $c \simeq 2 \cdot 10^4$ m/s for
orthoferrites.~\cite{BICh}

The formal Lorentz-invariance of spin dynamics of antiferromagnets
manifests itself in the motion of any AFM solitons,\cite{IBarIv}
in particular, the motion of the AFM vortex
core:~\cite{IvShekaPRL04,GaryMass} the dynamical equation for the
core coordinate at small vortex speed $\vec X$, when at $|d\vec X
/ dt| \ll c$, possess an inertial term,
 $$M_{\mathrm{v}}\frac{d^2\vec X}{dt^2}  = \vec F\,,$$
where the effective vortex mass $M_{\mathrm{v}} = E_{\mathrm{v}} /
c^2$, and  $\vec F$ is an external force acting on vortex. For the
case of interest here, the free dynamics of the vortex in a
circular sample, $\vec F$ is the restoring force: in  linear
approximation $\vec F = - \kappa \vec X$, where $\kappa $ is the
stiffness coefficient. With this force, the vortex core dynamics
is not a precession, as for the gyroscopic Thiele equation for
ferromagnetic vortices,~\cite{Thiele,Huber,Sonin} but rectilinear
oscillations, $\vec X(t) = \vec a\cos (\omega _{\mathrm{v}} t +
\phi _0 )$, degenerate with respect to the direction $\vec a$ and
$\phi _0 $, with  frequency $\omega _{\mathrm{v}} = \sqrt {\kappa
/ M_{\mathrm{v}}} $. For the easy-plane AFM model with $W_m = 0$,
such dynamics has been observed by direct numerical
simulations.~\cite{AFMvort91,AFMvort96} For a vortex state dot
with $R > R_{\mathrm{crit}} $, the value of $\kappa $ is
determined by the demagnetizing field, its value can be obtained
from the known value for a ferromagnet by replacing $M_s \to
M_{\mathrm{DM}}$,~\cite{GusIv} which gives $\kappa = 10\cdot 4 \pi
M^2_{\mathrm{DM}}L^2/9R$, and
\begin{equation} \label{v}
\omega _{\mathrm{v}} = \frac{2 cM_{\mathrm{DM}} \sqrt {10 L }}{3
\sqrt {A R \ln (\rho R / l_0 )} } \,.
\end{equation}

A simple estimate gives that $\omega _{\mathrm{v}} $, as for a
ferromagnetic vortex, is in the subGHz region, but with different
(approximately square root, instead of linear for ferromagnetic
vortex) dependence  on the aspect ratio $L / R$.

\subsection{Other low-frequency modes.}
Far from the vortex core, the other modes from this set  are
approximately characterized  by in-plane oscillations of $\vec l$
and $\vec M$. As their frequencies are small, $\omega \ll \gamma
H_{\mathrm{DM}} $, for these modes the magnetization $\vec {M}$ is
determined mainly by the in-plane static contribution \eqref{eq5},
and the formulae for the demagnetization field energy for
ferromagnetic vortices can be used. Moreover, the data known for
the ferromagnetic magnon frequencies $\omega^{\mathrm{FM}}_{m}$
(in first approximation over the small parameter $L/R$, i.e., in
the magnetostatic approximation), can be directly used for the
calculation of the corresponding frequencies for magnon modes
above the AFM vortex ground state.  Note that, in this
approximation,
$\omega^{\mathrm{FM}}_{+m}=\omega^{\mathrm{FM}}_{-m}=
\omega^{\mathrm{FM}}_{m}$, because the doublet splitting for a
ferromagnetic vortex is proportional to $(L/R)^2$; see
Refs.~\onlinecite{IvZasPRL05,GusSlav+Doubl}.

To make this essential simplification, note that for the
ferromagnetic case the Landau-Lifshitz equation, linearized over
the vortex ground state, can be written as $\partial m_r/\partial
t= 4 \pi \gamma M_s m_z$, $\partial m_z/\partial t= \hat h_2 M_s
m_r$, as was shown in Appendix B of Ref.~\onlinecite{BuessPRB05}.
Here the dimensionless operator $\hat h_2$ determines the
non-local magnetostatic part of the magnetic dipole interaction.
Then the equations can be easily rewritten as $\partial ^2
m_r/\partial t^2+ (\omega^{\mathrm{FM}}_{m})^2 m_r=0$, where
$\omega^{\mathrm{FM}}_{m}=4 \pi \gamma M_s \sqrt{\langle \hat h_2
\rangle}$, where $\langle \hat h_2 \rangle$ is the eigenvalue of
the operator $\hat h_2$. These values for modes with different
angular dependence can be either estimated theoretically or taken
from experiments.~\cite{BuessPRB05}

Now we will return to the case of AFM vortices.  Neglecting the
vortex core contribution, the second equation of the system
\eqref{eqLin} reduces to the form
 $$\partial ^2 \mu/\partial t^2 = (c/l_0)^2 \hat H_2 \mu \,,$$
 having exactly the same structure as the equation for $m_r$ for
a ferromagnetic vortex. For the simplified form of the
magnetization, $\vec M= M_{\mathrm{DM}}(\vec l \times \vec e_z) $
the same ``magnetic charges'', both volume charges, $\mathrm{div}
\vec M$, and surface charges, $(\vec e_r \cdot \vec M)$, as for a
ferromagnetic vortex, are produced. Thus the operator $\hat h_2$
differs from  $\hat H_2$ by a simple scaling relation, $ M_s \to
M_{\mathrm{DM}} $ and $A/4\pi M_s^2 \to l_0^2 $, which gives $\hat
H_2 = (4\pi M_{DM}^2/K )\hat h_2$. Then, it is easy to obtain the
frequency of the AFM mode $\omega _m$ in terms of  the frequency
of the ferromagnetic mode $\omega^{\mathrm{FM}}_{m}$ (if it is
known) as follows
\begin{equation} \label{omega-m}
\omega _{m} = \frac{c M_{\mathrm{DM}}}{\gamma M_s \sqrt{4\pi A}}
\cdot \omega^{\mathrm{FM}}_{m}.
\end{equation}

In particular, the frequency of radially symmetric oscillations,
having the highest frequency ~\cite{BuessPRB05} for modes with
minimal nodal number $n$, can be presented through the known value
for $\omega^{\mathrm{FM}}_{0}$ as~\cite{GalkinZasM0}
\begin{equation} \label{omega-m}
\omega _{0} = \frac{2 cM_{\mathrm{DM}} \sqrt { L }}{\sqrt {A R}}
\sqrt { \ln \left(\frac{6 R}{L}  \right) }  \,.
\end{equation}
For these modes, the frequencies are of the order of a few of GHz,
with an approximately square root dependence on the aspect ratio
$L / R$.

Note the absence of gyroscopic properties for the $\sigma$-model
equation. This is manifested both in the  absence of a gyroforce
for an AFM vortex, as well as in the fact that for an AFM vortex
the modes with the azimuthal numbers $m = \vert m\vert $ è $m = -
\vert m\vert $ are degenerate. The splitting of doublets with $m =
\pm \vert m\vert $, typical for  ferromagnetic vortices, is absent
for AFM vortices.~\cite{IvKolW,AFMvort96}

\subsection{Out of plane high-frequency modes.}
For an AFM sample in the vortex state, the high-frequency QAF
branch of magnons begets a set of discrete modes with frequencies
of the order of $\omega _\mathrm{g} $, i.e. hundreds of GHz. For
all these modes far from the vortex core, oscillations of the
vector $\vec {l}$ are out of plane, and the static contribution to
the weak magnetization, $\vec m_{\mathrm{static}} \propto (\vec
e_z\times \vec l)$ is absent, see \eqref{eq5}. Moreover, it is
easy to show that the dynamical part of $\vec {m} $, $\vec
m_{\mathrm{dyn}} \propto (\vec l \times \partial\vec l / \partial
t)$ has a vortex-like structure and does not lead to magnetic
poles neither on the up and down surfaces nor at the edge. Hence,
the dipole interaction is not essential for the description of
high-frequency magnons, and the results obtained earlier for the
vortex in easy-plane antiferromagnets without the magnetic dipole
interaction~\cite{IvKolW,AFMvort96} can be used as a good
approximation. In particular, the frequencies $ \omega _{n,m}$ are
close to $  \omega _\mathrm{g}$, and the difference $(\omega
_{n,m} -\omega _\mathrm{g}) $ decreases as the dot radius
increases.

For an AFM vortex within the local easy-plane model, the set of
high-frequency radially-symmetric modes with $m=0$ includes a
\emph{truly local} mode with an amplitude exponentially localized
within an area of the order of $5l_0 $ and with a frequency
$\omega _\mathrm{l} \sim 0.95\omega _\mathrm{g} $ independent on
$R$. Note that the frequency of this mode is \emph{inside} the
range of low-frequency in-plane modes, and hence, we have an
example of a truly local mode inside a continuum spectrum.

The presence of a local mode inside the frequency region of
low-frequency modes is quite a delicate feature, and it is
interesting to discuss whether or not such mode survives for the
vortex state AFM dots when accounting for dipole interactions. It
is easy to show that for these modes the oscillations of $\vec
{l}$ have no in-plane component, only the $\vartheta \neq 0$, even
inside the core region. Thus, for this mode the oscillations of
$\vec m$ are such that they do not disturb the vortex-like
closed-flux structure
 $$ \delta \vec m \propto \vec e_{\chi}[ ( \partial\vartheta / \partial t)
 +\gamma H_{\mathrm{D}}\vartheta \cos \theta _0 ]\,,$$
and all the magnetic poles vanish exactly. Hence, for cylindrical
dots made with canted antiferromagnets in the vortex state, a
radially symmetric mode with exponential localization inside an
area of radius 30--40 nm near the vortex core appears. The
frequency of this mode is approximately 5\% below the energy gap
of out-of-plane modes $\omega _\mathrm{g} $, which gives $\sim$ 9
GHz for hematite and $\sim$ 15 GHz for iron borate. We now stress
that such modes are absent for ferromagnetic vortices. This mode
can be imaged as an oscillation of the vortex core size, with
keeping the in-plane vortex-like structure for $\vec m$. The total
magnetic moment connected to this oscillations is zero, and the
excitation of these oscillations by a uniform magnetic field,
either pulsed or periodic, is impossible. However, such
oscillations can be excited by an instant change of the uniaxial
anisotropy, which determines the vortex core size. The novel
technique \cite{Kimel,kimel+09,Kalash,Satoh+} of spin excitations
by ultra-short laser pulses can be applied here, because the
linearly polarized light at inclined incidence, due to an
ultra-fast inverse Cotton-Mouton or inverse Voigt effect, is
equivalent to the necessary change of uniaxial anisotropy.

\section{Conclusion.}
To conclude, for micron-sized samples of typical canted
antiferromagnets, their ground state exhibits a topologically
non-trivial spin distribution. The magnetizations of each
sublattice $\vec {M}_1 $ and $\vec {M}_2 $ are characterized by a
vortex state with a standard out-of-plane structure, but the net
magnetization $\vec {M}=\vec {M}_1 +\vec {M}_2 $ forms a planar
vortex, where the projection of the magnetization normal to the
vortex plane everywhere in the sample is zero; in particular,
$\vec M=0$ in the vortex center. The vortex state AFM dots possess
a rich variety of normal magnon modes, from rectilinear
oscillations of the vortex core position with sub-GHz frequency to
out-of-plane modes with frequencies of the order of hundreds of
GHz, including a truly local mode. The use of QAF modes for vortex
state AFM dots, particularly the truly local mode, would allow the
application of magnonics for higher frequencies until $\sim$ 0.3
THz. This mode can be excited by ultra-short laser pulses with
linearly polarized light.

Our theory would be applicable to other systems with an AFM spin
structure, like a ferromagnetic bilayer dot containing two thin
ferromagnetic films with an AFM interaction between them,
described by the field $H_{\mathrm{ex}}$. If $H_{\mathrm{ex}}$ is
large enough, $H_{\mathrm{ex}} > 4 \pi M_s$, the anti-phase
oscillations of the magnetic moments of the layers produce
high-frequency modes with frequencies of the order of
$\sqrt{\gamma H_{\mathrm{ex}}\, \omega_{m,n}^{\mathrm{FM}}}$,
where $\omega_{m,n}^{\mathrm{FM}}$ are the frequencies of the
modes for a single layer dot.

\section*{Acknowledgments}

We gratefully acknowledge partial support from the  National
Security Agency, Laboratory of Physical Sciences, Army Research
Office, National Science Foundation grant No. 0726909, and
JSPS-RFBR contract No. 09-02-92114. E.G. and B.I. acknowledge
partial support from a joint grant from the Russian Foundation for
Basic Research and Ukraine Academy of Science, and from Ukraine
Academy of Science via Grant No. VC 38/ V 139-18.

\end{document}